# Wave packet dynamics in triplet states of Na$_2$ attached to helium nanodroplets

*Patrick Claas[1], Georg Droppelmann[1], Claus-Peter Schulz[2], Marcel Mudrich[3]\*, Frank Stienkemeier[3]*.

[1] Fakultät für Physik, Universität Bielefeld, 33615 Bielefeld, Germany

[2] Max-Born-Institut, Max-Born-Strasse 2a, 12489 Berlin, Germany

[3] Physikalisches Institut, Universität Freiburg, 79104 Freiburg, Germany

Mudrich@physik.uni-freiburg.de

The dynamics of vibrational wave packets excited in Na$_2$ dimers in the triplet ground and excited states is investigated by means of helium nanodroplet isolation (HENDI) combined with femtosecond pump-probe spectroscopy. Different pathways in the employed resonant multi-photon ionization scheme are identified. Within the precision of the method, the wave packet dynamics appears to be unperturbed by the helium droplet environment.

KEYWORDS (Word Style "BG_Keywords"). Femtosecond spectroscopy, helium nanodroplets, wave packet dynamics.

## Introduction

With the advent of femtosecond (fs) laser techniques it has become possible to directly observe nuclear dynamics of molecules and to chart the path of chemical reactions in real time [1-3]. In femtosecond pump-probe experiments of isolated molecules, a pump pulse prepares an initial wave packet, i.e. a coherent superposition of vibrational states which evolves in accordance with the time scales of vibrational (~$10^{-13}$ s) and rotational ($10^{-10}$ s) motion. The evolution of the wave packet is then probed with a time-delayed probe pulse via excitation to a specific final state which can be selectively detected. To date, fs pump-probe spectroscopy is an established technique with applications ranging from fundamental studies of the photodissociation dynamics of small molecules to the control of energy transfer in biological systems [4,5].

Real-time studies of molecules embedded in inert noble-gas matrices are motivated by a variety of new many-body effects prototypical for chemistry and biology in a condensed environment, such as solvational shifts of electronic states, electron-phonon coupling, matrix-induced electronic and vibrational relaxation and decoherence, reduced dissociation probability ('caging effect'), excimer and exciplex formation, charge recombination [6]. At the same time, these conceptually simple model systems are still accessible to theoretical treatment. The pioneering work by Zewail and coworkers using $I_2$ as a molecular probe and varying the structural properties of the solvent environment from gas, over cluster and liquid to solid, has been extended to a number of halogen molecules as well as $Hg_2$ isolated in rare gas matrices [7-11]. Due to the difficulty of implanting impurities into liquid helium, no real-time measurements with molecules in bulk liquid or solid helium have been reported until now. In a recent publication from our group, we presented for the first time pump-probe spectroscopy of covalently bound $K_2$ dimers attached to superfluid helium nanodroplets [12]. Different photoionization schemes were applied to study the influence of the helium environment on the wave packet propagation. Moreover, the desorption dynamics of $K_2$ off the helium droplets was directly observed.

In the present work, we report on pump-probe measurements of $Na_2$ dimers formed on helium nanodroplets in their triplet ground state. To our knowledge, this is the first time that wave packet dynamics has been observed in the weakly bound triplet manifold of alkali dimers. Due to the well-known potential curves, the easily accessible optical transitions and due to their simple preparation in molecular beams, alkali dimers are among the most thoroughly studied systems in the gas phase by means of fs spectroscopy. The wave packet dynamics of $Na_2$ in singlet states has been investigated in detail using multiphoton ionization by the group of G. Gerber [13-19]. Various ionization pathways were found depending on laser frequency and intensity [13,14]. At high laser intensities ($10^{12}$ W/cm$^2$), the formation of a wave packet in the electronic ground state $X^1\Sigma_g^+$ was observed [14]. Furthermore, interesting effects were studied such as revivals of the wave packet motion, wave packet propagation in spin-orbit coupled states, and the influence of wave packet dynamics on above-threshold ionization [15,19,20]. Photoelectron spectroscopy in combination with fs pump-probe laser pulses was found to be a sensitive method revealing a wealth of additional information not obtainable from the total yield of photoions [18,21,22]. Moreover, using phase shaped fs laser pulses, the coherent control of molecular multiphoton ionization was demonstrated [17,23]. Besides fs studies on Na$_2$ dimers, Na$_3$ trimers and Na$_n$ clusters have been studied in real time [24-26].

Helium nanodroplets are widely applied as a nearly ideal cryogenic matrix for spectroscopy of embedded molecules and as nanoscopic reactors for building specific molecular complexes [27-29]. Alkali atoms and molecules represent a particular class of dopant particles due to their extremely weak binding to helium droplets with binding energies on the order of 10 K (7 cm$^{-1}$). From both theory and experiment it is known that alkali dimers reside in bubble-like structures on the surface of helium droplets [30,31]. Therefore, spectroscopic shifts of the electronic excitation spectra of alkali dimers are in the range of only a few cm$^{-1}$ with respect to the gas phase [27,28,31,32].

Collisional, internal as well as binding energies are dissipated by the helium droplet through evaporation of helium atoms, which may cause desorption of the alkali dimers from the droplets. Since the amount of internal energy released upon formation of ground state ($X^1\Sigma_g^+$) dimers greatly exceeds the one released upon formation of dimers in the lowest triplet state $a^3\Sigma_u^+$, the latter have a higher chance to remain attached to the droplets. This leads to an enrichment of the droplet beam with high-spin dimers and clusters compared to covalently bound systems [31-36]. Thus, HENDI opens the possibility of spectroscopically studying the triplet manifold of alkali dimers which is difficult to reach by standard techniques [31-34]. The alkali-helium droplet complex eventually equilibrates at the terminal temperature of pure helium droplets of 380 mK. Thus, only the lowest vibrational state $v=0$ and a few rotational states are populated which provides well-defined starting conditions for a pump-probe experiment. In contrast to conventional molecular beam experiments, this is true even for the weakly bound triplet ground state $a^3\Sigma_u^+$ with a vibrational constant in the range of 20 cm$^{-1}$. Upon electronic excitation, alkali atoms and molecules mostly desorb from the droplets as a consequence of evaporation of helium atoms after energy deposition in the helium droplet. The dynamics of the desorption process has been observed for the first time by analyzing the wave packet motion of K$_2$ dimers on helium droplets [12].

Due to their weak coupling to the surrounding helium environment, alkali dimers on helium droplets can be viewed as an intermediate system between free molecules and molecules isolated in conventional cryogenic matrices. The wave packet propagation is expected to be only weakly perturbed. The highly quantum nature of superfluid $^4$He droplets vs. normal fluid $^3$He droplets may affect the coupling of the wave packet motion to the droplet. Moreover, exotic high-spin molecules and complexes formed on He droplets have been studied in real time [31-36].

**Experimental**

The experimental setup is described in detail in Ref. 12 and is only briefly discussed here. It comprises three main parts: a molecular beam apparatus that provides Na$_2$-doped helium droplets, a laser system to produce pairs of fs laser pulses and a mass selective ion detector. The beam of helium droplets of mean size around 5000 helium atoms is produced by expanding high purity helium into vacuum through a nozzle 5 μm in diameter at a stagnation pressure of 80 bar and at a nozzle temperature of about 23 K. Doping of helium droplets with Na$_2$ molecules in the $v$=0 triplet ground state is achieved by successive pick up of two Na atoms in a vapor cell. The number of collisions of the droplets with free dopant atoms inside the vapor cell is controlled by its temperature. Given the flight distance of the droplets of 1 cm inside the vapor cell a temperature T≈485 K ($p_{Na}$ ≈ 4 10$^{-4}$ mbar) of the Na reservoir provides the highest probability for pick-up of two dopant atoms per droplet.

The beam of doped helium droplets is intersected perpendicularly by the laser beam inside the detection volume of a quadrupole mass analyzer. Thus, Na$_2^+$ dimer ions which have been optically excited and ionized by absorption of three or more laser photons and which have desorbed from the helium droplets are detected mass selectively. The laser pulses having an average output power of around 1.2 W are generated by a Ti:sapphire laser at 80 MHz repetition rate. The pulse width is about 110 fs corresponding to a spectral width of 135 cm$^{-1}$ (FWHM). The laser pulses are split into pairs with variable delay using a Mach-Zehnder interferometer.

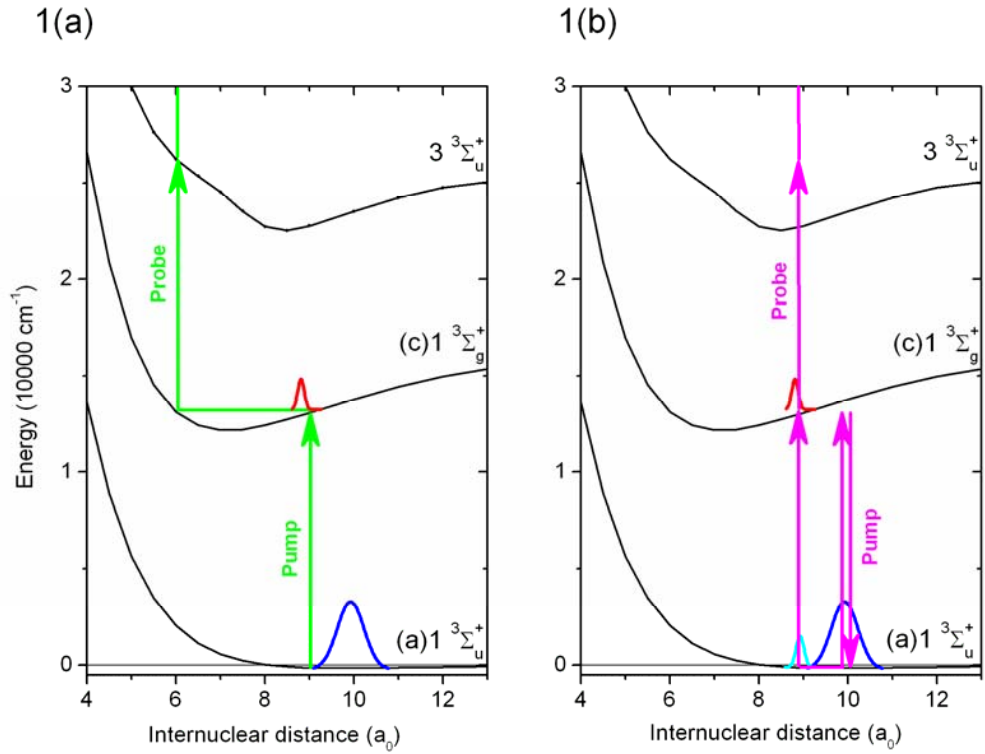

**Figure 1.** Potential energy curves of Na$_2$ populated in the fs pump-probe experiment [37]. The arrows indicate pump and probe transitions leading to photoionization according to two different pathways: One-photon excitation of a wave packet in the *c*-state followed by resonant two-photon ionization (a) and creation of a wave packet in the initial *a*-state by resonant impulsive Raman scattering followed by three-photon ionization (b). Only selected relevant curves of the triplet manifold are included.

**Results**

The one-color pump-probe experiments reported in this paper are carried out in the wave length region λ=733-761 nm (13140-13640 cm$^{-1}$). The excitation scheme leading to wave packet propagation and subsequent ionization is illustrated in figure 1 using the unperturbed potential curves from Ref. 37. The relevant curves are the triplet ground state $a^3\Sigma_u^+$, the first excited triplet state $c^3\Sigma_g^+$, and the excited state $3^3\Sigma_u^+$. The ionic continuum to which the Na$_2$ dimers are eventually ionized is not shown. The arrows in figure 1(a) and (b) indicate two different excitation pathways leading to photoionization which are identified in this work. In both cases the excitation starts from the vibrational ground state $a^3\Sigma_u^+$, $v=0$. In the case shown in (figure 1(a)), the absorption of one photon from the pump pulse creates a wave packet by coherent superposition of about 3 vibrational states around $v'=10$ in the $c$-state close to the classical outer turning point [32]. The wave packet then propagates inwards. At the inner turning point the probe pulse can ionize the excited Na$_2$ in a two-photon step efficiently since the photon energy comes into resonance with the energy difference between the $c^3\Sigma_g^+$ and $3^3\Sigma_u^+$ states. Thus, a Franck-Condon (FC) window for the resonance-enhanced ionization opens at a well-defined internuclear distance and therefore the wave packet dynamics in the $c$-state is filtered out. The measured photoionization yield shows an oscillatory structure corresponding to the vibrational motion of the molecules (cf. figure 2). Since the time between the creation of the wave packet and the first appearance of the Na$_2$ ions roughly equals half the period of the wave packet oscillation the measured oscillation has a well-defined phase $\Phi \approx \pi$ at zero delay times. In the second case (figure 1(b)), a wave packet is created in the initial $a^3\Sigma_u^+$ electronic state by resonant impulsive Raman scattering (RISRS) through coherent superposition of about 3 vibrational states around $v=1$. This wave packet, initially centered around the minimum of the potential well, propagates towards the inner turning point and is ionized by three-photon absorption. Therefore, the time delay between wave packet creation and ionization approximately equals one quarter of the oscillation period such that one observes a phase at zero delay time $\Phi \approx \pi/2$. Which one of the two photoionization schemes dominates the ion signal depends on the laser wave length and the pulse energy [13,14].

The raw pump-probe transients of Na$_2$ attached to helium nanodroplets are processed by applying a band-pass filter which suppresses slowly varying contributions to the ion count rate as well as high-frequency noise, as discussed in detail in Ref. 12. The time range 0-600 fs is omitted because the signal is entirely masked by the autocorrelation trace of the copropagating pump and probe pulses. Typical processed real-time pump-probe transients at different wave lengths of the femtosecond laser are displayed in figure 2.

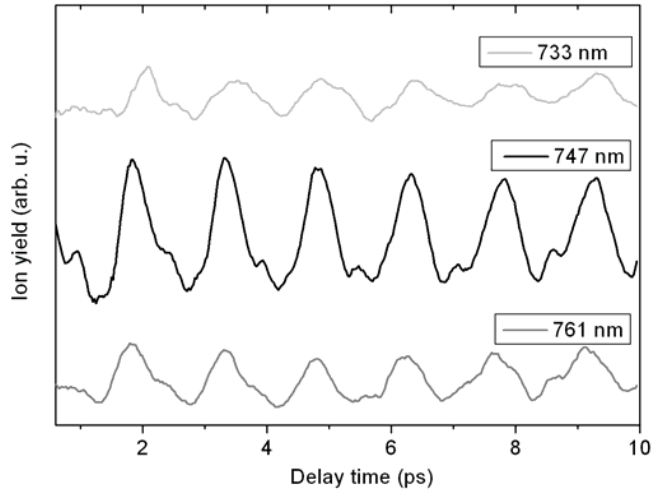

**Figure 2**. Pump-probe transients of $Na_2$ on helium nanodroplets at different laser wave lengths.

The best contrast of oscillations in the photoionization signal is achieved at the laser wave length $\lambda=747$ nm. The signal is dominated by an oscillation with a period of about 1.5 ps. An additional superimposed periodic signal is visible as small peaks in between the dominating maxima. At wave lengths $\lambda=761$ nm and $\lambda=733$ nm the transient signals have weaker contrast and are clearly dominated by a strong oscillation of the same period as observed at $\lambda=747$ nm. In a first step the data are analyzed in the time domain by assigning index numbers to the most prominent maxima and by performing linear regression of the index numbers as a function of delay time [12]. This yields the oscillation periods $T_{747nm}=1484(2)$ fs, $T_{761nm}=1470(6)$ fs, and $T_{733nm}=1468(5)$ fs and phases $\Phi_{747nm}=0.496(3)\pi$ (368(2) fs), $\Phi_{761nm}=0.52(3)\pi$ (383(20) fs), and $\Phi_{733nm}=0.78(4)\pi$ (573(26) fs). The numbers in parentheses indicate the experimental uncertainty of the last digits. At $\lambda=747$ nm and $\lambda=761$ nm, the phase lag roughly equals $\Phi=\pi/2$ indicating that the oscillation is dominated by wave packet motion according to the second scheme described above, in which a wave packet is created in the triplet ground state by RISRS. The additional weaker oscillation at λ=747 nm can be isolated by applying a narrow band pass filter around the frequency 88 cm$^{-1}$ (T=380(4) fs) to the data and by repeating the analysis procedure described above. The resulting phase is Φ'$_{747nm}$ =1.00(1)π. Thus, this weak oscillation can be attributed to the dynamics of a wave packet in the potential of the first excited state $c^3\Sigma^+_g$.

In order to obtain more quantitative information about the exact frequencies and relative amplitudes of the contributing frequency components the pump-probe traces are Fourier transformed in the entire recorded delay time interval. The resulting spectrum obtained from the scan recorded at λ=747 nm is shown in Figure 3. The spectrum comprises 3 groups of frequencies. The frequencies of the first most prominent group are 18.1(6) cm$^{-1}$, 20.7(7) cm$^{-1}$, and 23.3(5) cm$^{-1}$. These frequencies constitute the dominant wave packet oscillation in the triplet ground state $a^3\Sigma^+_u$ shown in figure 2 and correspond to the vibrational level spacing *ΔG(v,v+1)=G(v)-G(v+1)* between neighboring vibrational levels *v, v+1*. Table 1 compares the measured values of this work (*ΔG$_{HENDI}$*) with experimental values obtained from fluorescence spectroscopy in the gas-phase by Li *et al* [38] (*ΔG$_{exp}$*), and theoretical results by Ho *et al*. [39](*ΔG$_{theo}$*). This comparison shows clearly, that the observed frequency components 1-3 can be assigned to level spacing *ΔG(0,1), ΔG(1,2),* and *ΔG(2,3)*, respectively. Frequency component 4 (45.3(6) cm$^{-1}$) matches within the error bar twice the value *ΔG(0,1)*. Therefore it can be attributed to coherent excitation of vibrational levels with *Δv=2*, in particular *ΔG(0,2),* and *ΔG(1,3)*. These results in combination with the fitted phase Φ≈π/2 confirm the interpretation in terms of wave packet dynamics induced in the triplet ground state $a^3\Sigma^+_u$ by RISRS, followed by three-photon ionization at the inner turning point of the *a*-state potential.

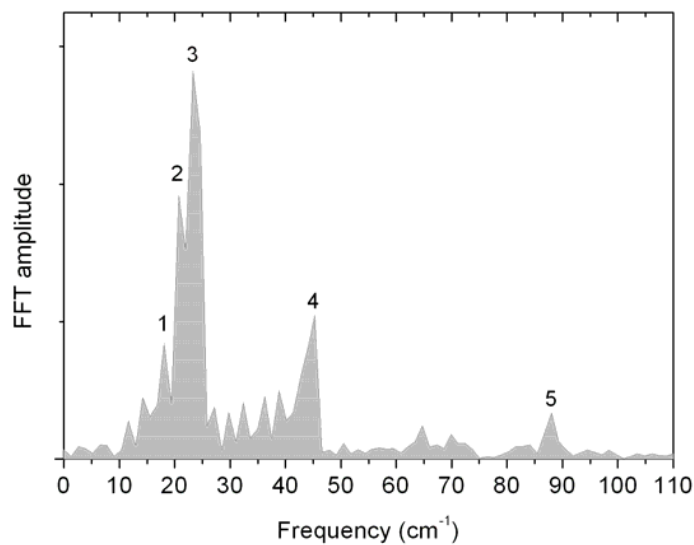

**Figure 3**. Fourier transform of the pump-probe scan at laser wave length $\lambda$=747 nm. The frequency components labeled with numbers 1-4 reflect the dynamics of wave packets in the *a*-state (solid arrows in figure 1) and the component number 5 reflects the wave packet dynamics in the *c*-state (dashed arrows in figure 1).

In order to identify frequency component number 5 (88.0(6) cm$^{-1}$) it is compared to the values from fluorescence spectroscopy of Na$_2$ attached to helium droplets [32]. Accordingly, at $\lambda$=747 nm the vibrational states *v'*=9-11 of the first excited state $c^3\Sigma_g^+$ are populated when taking into account the spectral width of the laser pulses. The vibrational level spacing, $\Delta G(9,10)$= 88.1 cm$^{-1}$ and $\Delta G(10,11)$= 87.2 cm$^{-1}$ are in good agreement with the value found in this work within the experimental uncertainty. Due to the limited spectral resolution of the Fourier spectra, the two frequencies are not resolved and presumably both contribute to the observed signal. This agreement in connection with the fitted phase of the oscillation $\Phi'_{747nm} \approx \pi$ implies a pump-probe ionization scheme in which the wave packet motion is induced in the $c^3\Sigma_g^+$ state by the pump pulse and probed by the probe pulse at the opposite, inner turning point, as previously described.

| (v,v+1) | $\Delta G_{exp}$ (cm$^{-1}$) | $\Delta G_{theo}$ (cm$^{-1}$) | $\Delta G_{HENDI}$ (cm$^{-1}$) |
|---|---|---|---|
| (0,1) | 23.67 | 22.94 | 23.3 |
| (1,2) | 20.97 | 21.32 | 20.7 |
| (2,3) | 19.59 | 19.62 | 18.3 |

**Table 1**. Comparison of energy spacings between vibrational levels $v$ in the triplet ground state $a^3\Sigma_u^+$ obtained from gas-phase spectroscopy ($\Delta G_{exp}$) [38], theory ($\Delta G_{theo}$) [39], and from this work ($\Delta G_{HENDI}$).

Fourier analysis of the pump-probe transients at λ=733 nm and λ=761 nm essentially reveals only one frequency component at about 20 cm$^{-1}$ which is attributed to the dynamics in the triplet ground state $a^3\Sigma_u^+$, in analogy with the discussion above. The missing frequency component corresponding to the dynamics in the $c^3\Sigma_g^+$ state is probably due to the weak contrast of the signal. Alternatively, the creation or the detection of such a wave packet may be suppressed.

**Discussion**

The presented measurements reveal that vibrational wave packets can readily be placed into the triplet ground and the excited *c*-state. The corresponding Franck-Condon windows project the wave packet motion with high contrast to the final ionic state. Evaluation of the phases allows to locate both the radial position when forming the wave packet as well as the radial position of the Franck-Condon detection window. In this regard the RISRS process comes out as a peculiar case where the wave packet appears to be formed at the center of the ground state potential well, while the ionization step clearly is very much enhanced at the inner turning point.

Vibrational shifts of molecules attached to helium nanodroplets typically lie in the range 0.1-2 cm$^{-1}$ (Ref. 27). Since alkali dimers reside on the surface, even smaller shifts might be expected. The results reported here give vibrational energies that go along with the gas phase values within the errors of the experiments. Accordingly, earlier measurements with K$_2$ on helium droplets revealed helium induced shifts of level spacings with respect to the gas phase of less than 1 cm$^{-1}$ (Ref. 12).

At all three laser wave lengths, the dominating channel is the multi-photon ionization according to the scheme involving wave packet motion in the triplet ground state $a^3\Sigma^+_u$ induced by RISRS. The wave packets induced in the $c^3\Sigma^+_g$ state give significantly less contrast in the pump-probe transients. This can be interpreted qualitatively using the following arguments. At $\lambda=747$ nm, the excitation of a wave packet directly in the $c$-state is considerably suppressed by Franck-Condon principle as indicated by the poor overlap of the initial wave function $v=0$ in the $a^3\Sigma^+_u$ state with the wave packet in the $c$-state shown in figure 1(a). The change in the radial coordinate followed from the wave packet motion in the $c$-state, however, opens a resonant window to the ionic continuum for the probe step at the inner turning points of the states $c^3\Sigma^+_g$ and $3^3\Sigma^+_u$.

In contrast, the wave packet induced in the $a$-state by RISRS has at the inner turning point large overlap with vibrational states in the potential of the $c$-state, as illustrated in figure 1(b). Thus, the probe transition to the ionic continuum is resonantly enhanced. In default of potential curves at higher excitation energies, it cannot be determined whether the second stage involved in the probe step is also resonant. However, since the density of electronic states increases with energy, it is likely that the second stage before reaching the ionic continuum is resonance enhanced as well. In total, this second pathway to ionization turns out to be more efficient, even though a two-photon transition is involved. Other potential curves have been considered to interpret the observed wave packet dynamics but none were found at these particular excitation energies having appropriate symmetries. The same arguments can be applied to the observations at laser wave length $\lambda=733$ nm and $\lambda=761$ nm. At $\lambda=733$ nm, transition probability from $a^3\Sigma^+_u$, $v=0$ to the $c$-state is higher than at $\lambda=747$ nm, however, the probe step to the $3^3\Sigma^+_u$ state is no longer resonant. At $\lambda=761$ nm, transition probability to the $c$-state is even weaker than at $\lambda=747$ nm.

In the dominating RISRS process, the evolution of the vibrational wave packet in the ground state drives the internuclear distance into resonant conditions for ionization at the inner turning point. Since the vibrational states of the triplet ground state could readily be populated by thermal energies in normal gas phase experiments, the process observed in our experiment underlines the importance of the low temperatures, having the ensemble of molecules only in the vibrational ground state. A mixture of populated vibrations from a thermal distribution would probably wash out FC windows and hamper the observation of wave packet motions because of direct resonant conditions from the electronic ground state.

Earlier measurements with $K_2$ on helium droplets clearly revealed additional dynamics due to the desorption of $K_2$ off the helium droplets [12]. At certain excitation energies, both the oscillation frequencies and the Franck-Condon windows determining the relative amplitudes of different ionization pathways displayed a time dependent variation. In contrast to that, no such dynamics is observed with $Na_2$. Both oscillations in the *a* and in the *c*-states have nearly constant amplitudes and frequencies in the studied time window between 0 and 10 ps. Therefore, either the time scale for desorption of $Na_2$ greatly exceeds the pump-probe delay times studied in this work. Alternatively, the perturbation of the dynamics of $Na_2$ by the helium environment is so weak that the desorption process remains hidden.

**Conclusions**

In conclusion, the dynamics of vibrational wave packets in triplet states of $Na_2$ on helium nanodroplets is studied by resonant multi-photon ionization at different laser wave lengths. To our knowledge, this is the first demonstration of wave packet propagation in weakly bound triplet states. Two distinct ionization schemes are identified by analyzing frequencies and phases of the wave packet oscillations. The measured vibrational frequencies coincide with the gas-phase values within the error bars. This reflects the fact that helium nanodroplets only weakly perturb the vibrational structure of embedded molecules and therefore are very well suited as matrix for spectroscopy. Clearly, the low temperature conditions, in particular the presence of only vibrational ground state population, appears to be essential for the observed wave packet dynamics. In contrast to real-time experiments on potassium dimers, no dynamics to be attributed to the desorption of $Na_2$ or energy dissipation to the droplet could be extracted from the observed oscillations.

In the future, these measurements will be extended to longer pump-probe delay times in order to probe the dynamics on a longer time scale and to improve the frequency resolution of the Fourier spectra. In particular, the effects resulting from the perturbation due to the helium environment may be observable. Also, quantum dynamics simulations should be performed, introducing perturbed $Na_2$ potential curves to quantify the influence of the helium matrix.

ACKNOWLEDGMENT (Word Style "TD_Acknowledgments"). Technical support with the laser system from V. Petrov is gratefully acknowledged. The work is financially supported by the DFG.